\documentclass[twocolumn,preprintnumbers,amsmath,amssymb]{revtex4}

\usepackage{latexsym}%
\usepackage{graphicx}
\usepackage{dcolumn}
\usepackage{bm}

\begin{document}

\newcommand{\bea}{\begin{eqnarray}}
\newcommand{\eea}{\end{eqnarray}}
\newcommand{\be}{\begin{equation}}
\newcommand{\ee}{\end{equation}}

%
%
\def\shiftleft#1{#1\llap{#1\hskip 0.04em}}
\def\shiftdown#1{#1\llap{\lower.04ex\hbox{#1}}}
\def\thick#1{\shiftdown{\shiftleft{#1}}}
\def\b#1{\thick{\hbox{$#1$}}}

\title{Three-quark currents and baryon spin}

\author{A. J. Buchmann}
\email{alfons.buchmann@uni-tuebingen.de}
\affiliation{Institut f\"ur Theoretische Physik, Universit\"at T\"ubingen
Auf der Morgenstelle 14, D-72076 T\"ubingen, Germany}
\author{E. M. Henley}
\email{henley@u.washington.edu}
\affiliation{Department of Physics and Institute for Nuclear Theory, 
Box 351560, University of Washington, Seattle, WA 98195, USA}

\begin{abstract}
We show that three-quark axial 
currents as required by broken SU(6) spin-flavor symmetry 
reduce the quark spin contribution to proton spin 
from $\Sigma_p = 1$ (one-quark axial current value) to 
$\Sigma_p = 0.41(12)$
consistent with the empirical value $\Sigma_{p, exp} = 0.33(08)$.
In the case of the $\Delta^+(1232)$ baryon, we find that three-quark 
axial currents increase the one-quark axial current value
$\Sigma_{\Delta^+} = 3$ to $\Sigma_{\Delta^+} = 3.87(22)$.
We also calculate the quark orbital angular momenta $L_u$ and $L_d$ 
in the proton and $\Delta^+$ and interpret our results in terms of the
prolate and oblate geometric shapes of these baryons consistent with
their intrinsic quadrupole moments.
\end{abstract}

\maketitle

\section{Introduction}
\label{intro}
The question how the proton spin is made up from
the quark spin $\Sigma$, quark orbital angular momentum $L_q$,
gluon spin $S_g$, and gluon orbital angular momentum $L_g$
\be
\label{angmom}
J = \frac{1}{2} \Sigma + L_q + S_g + L_g
\ee
is one of the central issues in nucleon structure physics~\cite{seh74,ji97}.
In the constituent quark model with only one-quark operators, 
also called additive quark model,
one obtains $J=\Sigma/2 =1/2$, i.e., the proton spin is the sum
of the constituent quark spins and nothing else.
However, experimentally it is known that only about 1/3 of  the proton
spin comes from quarks~\cite{aid12}. The disagreement between the 
additive quark model
result and experiment came as a surprise because the same model
accurately described the related proton and neutron magnetic moments.
We show that the failure of the additive quark model to describe
the quark contribution to proton spin correctly is due to its neglect
of three-quark terms in the axial current~\cite{Hen11}.

\section{Broken spin-flavor symmetry and QCD parametrization method}
\label{sec:2}

In the present work, the general QCD parametrization method 
developed and explained in detail
by Morpurgo~\cite{Mor89} is used to calculate the quark contribution
to baryon spin in a systematic manner.
Previously, we have applied this method 
to calculate higher order corrections to
baryon-meson couplings~\cite{Hen00} and baryon
electromagnetic moments~\cite{Hen08}. Here, we construct the most 
general expression
for the quark angular momentum operator ${{\tilde \Omega}}$ 
in spin-flavor space that is
compatible with the space-time and inner QCD symmetries.

The first step is to realize~\cite{Sak64} that a general SU(6) spin-flavor 
operator ${\tilde \Omega}^{R}$ acting on the ${\bf 56}$ 
dimensional baryon ground state
supermultiplet must transform according to one of the irreducible
representations $R$ contained in the direct product
$\label{directproduct}
\bar{{\bf 56}} \times {\bf 56}
=  {\bf 1} + {\bf 35} + {\bf 405} + {\bf 2695}.$ 
The ${\bf 1}$ dimensional representation (rep) 
corresponds to an SU(6) symmetric operator,
while the ${\bf 35}$, ${\bf 405}$,
and ${\bf 2695}$ dimensional reps characterize respectively, first, second,
and third order SU(6) symmetry breaking. Therefore, a general SU(6) symmetry 
breaking operator for ground state baryons has the form
\be
\label{genop}
{\tilde \Omega}
= {\tilde \Omega}^{\bf 35} +
{\tilde \Omega}^{\bf 405} + {\tilde \Omega}^{\bf 2695}.
\ee

The second step is to decompose each SU(6) tensor ${\tilde \Omega}^R$ 
in Eq.(\ref{genop}) 
into SU(3)$_F\times$SU(2)$_J$ subtensors ${\tilde \Omega}^R_{(F,2J+1)}$,
where $F$ and $2J+1$ are the dimensionalities of the flavor and spin reps. 
One finds ~\cite{Hen11,Beg64a} that a flavor singlet $(F=1)$ axial 
vector $(J=1)$ operator
${\tilde \Omega}^{R}_{({\bf 1}, {\bf 3})}$ 
needed to describe baryon spin, 
is contained {\it only} in the $R={\bf 35}$ and $R={\bf 2695}$ 
dimensional reps of SU(6). 

The third step is to construct quark operators transforming 
as the SU(6) tensor  
${\tilde \Omega}^R_{({\bf 1},{\bf 3})}$.
In terms of quarks, the SU(6) tensors on the right-hand side of
Eq.(\ref{genop}) are represented respectively by one-, two-, 
and three-quark operators~\cite{leb95}. We find the following 
uniquely determined 
one-quark ${\bf A}_{[1]}$ and three-quark ${\bf A}_{[3]}$ 
flavor singlet axial currents~\cite{Hen11} 
\bea
{\tilde \Omega}^{\bf 35}_{({\bf 1},{\bf 3})} & = &
{\bf A}_{[1]}  =  A \, \sum_{i=1}^3 \  {\b{\sigma}}_{i}, \nonumber \\
{\tilde \Omega}^{\bf 2695}_{({\bf 1},{\bf 3})}  &=& 
{\bf A}_{[3]}  =   C \, \sum_{i \ne j \ne k}^3  
\ {\b{\sigma}}_i \cdot {\b{\sigma}}_j \ {\b{\sigma}}_{k},
\eea 
where $\b{\sigma}_i$ is the Pauli spin matrix of quark $i$. The constants
$A$ and $C$ are to be determined from experiment.
The most general flavor singlet axial current compatible with broken SU(6) 
symmetry is then 
\be 
\label{total}
{\bf A} = {\bf {A}}_{[1]} + {\bf {A}}_{[3]}
= A \, \sum_{i=1}^3 \  {\b{\sigma}}_{i} +
C \, \sum_{i \ne j \ne k}^3  \ {\b{\sigma}}_i \cdot {\b{\sigma}}_j
\ {\b{\sigma}}_{k}.
\ee
The additive quark model corresponds to $C=0$ and $A=1$. 
The three-quark operators are an effective description of quark-antiquark 
and gluon degrees of freedom. 

\section{Quark spin contribution to baryon spin}
\label{sec:3}

By sandwiching the flavor singlet axial current ${\bf A}$ of Eq.(\ref{total})
between standard SU(6) baryon wave functions~\cite{lic70} we obtain 
for the quark spin contribution to the spin of octet and 
decuplet baryons~\cite{Hen11} 
\bea
\label{matrixelements}
\Sigma_{1}:  & = &
\langle B_8 \uparrow \vert {\bf A}_z \vert  B_8 \uparrow \rangle = A - 10\, C,
\nonumber \\
\Sigma_{3}:  & = &  
\langle B_{10} \uparrow \vert {\bf A}_z \vert  B_{10} \uparrow \rangle =
3\,A + 6\,C,
\eea
where $B_8$ ($B_{10}$) stands for any member of the
baryon flavor octet (decuplet). Here, $\Sigma_1$ ($\Sigma_3$) 
is twice the quark spin
contribution to octet (decuplet) baryon spin. Our theory predicts 
the same quark contribution 
to baryon spin for all members of a given flavor multiplet, because
 the operator in
Eq.(\ref{total}) is by construction a flavor singlet that does not break SU(3) 
flavor symmetry. 
On the other hand,  SU(6) spin-flavor symmetry is broken as reflected by the 
different expressions 
for flavor octet and decuplet baryons. 

To proceed, we construct from the operators in Eq.(\ref{total}) 
one-body ${\bf A}^q_{[1] \, z}$ and three-body ${\bf A}^q_{[3] \, z}$ 
operators of flavor $q$ acting only on $u$ quarks and $d$ quarks~\cite{Hen11}  
\bea
\label{u-quark1and3b}
{\bf A}_z^u
& = & A \sum_{i=1}^{3} \b{\sigma}^u_{i\, z}
+2 C \sum_{i \ne j \ne k}^{3} 
{\b{\sigma}}^u_i \cdot {\b{\sigma}}^d_j\ {\b{\sigma}}^u_{k\, z}, \nonumber \\
{\bf A}_z^d
& = & A \sum_{i=1}^{3} \b{\sigma}^d_{i\, z} +
C \sum_{i \ne j \ne k}^{3} 
{\b{\sigma}}^u_i \cdot {\b{\sigma}}^u_j\ {\b{\sigma}}^d_{k\, z}.
\eea  
For the $u$ and $d$ quark contributions to the spin of the proton we obtain
\bea
\label{flavordecomp}
\Delta u & = &
\langle p \uparrow \vert \,
{\bf A}_{[1]\, z}^u + {\bf A}_{[3]\, z}^u  \vert  p \uparrow \rangle
 =  \phantom{-}\frac{4}{3}\, A  - \frac{28}{3} \, C, \nonumber \\
\Delta d  & = &
\langle p \uparrow \vert \,
{\bf A}_{[1]\, z}^d + {\bf A}_{[3]\, z}^d  \vert  p \uparrow
\rangle
 =  -\frac{1}{3}\, A  - \frac{2}{3} \, C.
\eea
These theoretical results are to be compared with the combined deep inelastic
scattering and hyperon $\beta$-decay experimental data, from
which the following quark spin contributions to the
proton spin were extracted~\cite{aid12} 
$\Delta u = \hspace{.35cm} 0.84 \pm 0.03,\hspace{.5cm}
\Delta d =-0.43 \pm 0.03, \hspace{.5cm}
\Delta s = -0.08 \pm 0.03.$
The sum of these spin fractions
$\Sigma_{1_{exp}}=\Delta u + \Delta d + \Delta s = 0.33(08)$ is considerably
smaller than expected from the additive quark model, which gives $\Sigma_1=1$.

Solving Eq.(\ref{flavordecomp}) for $A$ and $C$ fixes 
the constants $A$ and $C$ as
\be
A  =  \phantom{-}\frac{1}{6}\, \, \Delta u  - \frac{7}{3}\, \Delta d,
\qquad 
C  =  -\frac{1}{12}\, \Delta u - \frac{1}{3}\, \Delta d.
\ee
Inserting the experimental results for $\Delta u$ and $\Delta d$
we obtain $A=1.143(70)$ and $C=0.073(10)$ and from Eq.(\ref{matrixelements}) 
\bea
\Sigma_{1} &  = & A - 10\, C = 1.14 - 0.73 = 0.41(12), \nonumber \\
\Sigma_{3} &  = &3\,A + 6\,C =3.42 + 0.45 = 3.87(22)
\eea
compared to the experimental result $\Sigma_{1_{exp}}= 0.33(08)$.
For octet baryons, the three-quark term is of the same importance
as the one-quark term because of the factor 10 multiplying $C$.
It is interesting that for decuplet baryons, quark spins
add up to 1.3 times the additive quark model value  $\Sigma_3 = 3$.

\section{Quark orbital angular momentum contribution to baryon spin}
\label{sec:4}
In this section we apply the spin-flavor operator analysis 
of Sect.~\ref{sec:3}) to quark orbital angular momentum $L_z$ 
using the general operator 
of Eq.(\ref{total}) for $L_z$ with new constants $A'$ and $C'$
\bea
\label{matrixelements2}
L_z(8) &  = &   
\langle B_8 \uparrow \vert { L}_z \vert  B_8 \uparrow \rangle 
= \frac{1}{2} \left ( A' - 10\, C' \right ), \nonumber \\
L_z(10)  &  = & 
\langle B_{10} \uparrow \vert {L}_z \vert  B_{10} \uparrow \rangle 
= \frac{1}{2} \left ( 3\,A' + 6\,C' \right ).
\eea
Assuming that the gluon total angular momentum 
$S_g+L_g \approx 0$ is small~\cite{aid12} 
we obtain from Eq.(\ref{angmom})
\bea
\label{matrixelements3}
L_z(8) & = &  \frac{1}{2} - \frac{1}{2} \Sigma_1  =  0.30,  \nonumber \\
L_z(10)& = &  \frac{3}{2} - \frac{1}{2} \Sigma_3 = -0.44. 
\eea
Eq.(\ref{matrixelements2}) and Eq.(\ref{matrixelements3}) yield for the 
parameters $A'=1-A=-0.143$
and $C'=-C=-0.073$. 
Next, we calculate the orbital angular momentum carried by $u$ and
$d$ quarks in the proton in analogy to Eq.(\ref{flavordecomp})
\bea
\label{flavordecomp_orb_proton}
L_z^u(p)& = &  \frac{1}{2} \left (
\frac{4}{3}\, A'  - \frac{28}{3} \, C' \right )=0.25, 
\nonumber \\
L_z^d(p)  &  = & \frac{1}{2} \left ( -\frac{1}{3}\, A'  - \frac{2}{3} \, C' 
\right )=0.05.
\eea 
For the total angular momentum carried by quarks we get 
$J^u(p)=\frac{1}{2}\Delta u + L_z^u(p)=0.42+0.25=0.67$ and
$J^d(p)=\frac{1}{2}\Delta d + L_z^d(p)=-0.22+0.05=-0.17$.
Our results for $J^u(p)$ and $J^d(p)$ 
are consistent with those of Thomas~\cite{tho09} 
who finds $J^u(p)=0.67$ and $J^d(p)=-0.17$ at the low energy (model) scale.
Applying the $u$ and $d$ quark operators in Eq.(\ref{u-quark1and3b}) to the 
$\Delta^+$ state we obtain
\bea
\label{flavordecomp_orb_Delta}
L_z^u(\Delta^+)& = & \frac{1}{2} \left ( 2 \, A'  + 4 \, C' \right )=-0.29,
\nonumber \\
L_z^d(\Delta^+)& = & \frac{1}{2} \left ( A'  + 2 \, C' \right )=-0.15.
\eea

\begin{widetext}
\begin{figure*}
\resizebox{0.80\textwidth}{!}{
\includegraphics{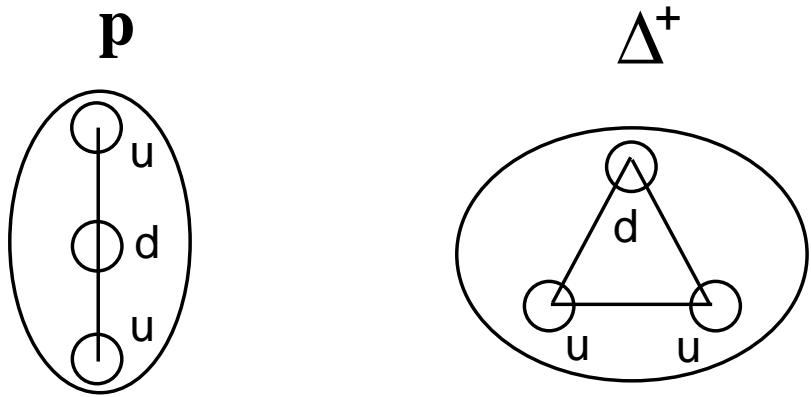}}
\vspace{-1.2cm}
\caption{
Qualitative picture of the $u$ and $d$ quark distributions in the 
proton (left) and $\Delta^+$ (right).
In the proton, most of the 
quark orbital angular momentum is carried by the $u$ quarks 
and relatively little by the $d$ quarks.
This is consistent with a linear (prolate or cigar-shaped) 
quark distribution with 
the $u$ quarks at the periphery and the $d$ quark near the origin.
In contrast, in the $\Delta^+$, the $u$ quark orbital angular momentum 
is just twice that of the $d$ quark. This is 
consistent with a planar (oblate or pancake-shaped) quark distribution, 
in which each quark has the same distance from the origin.} 
\label{figure:shapes}
\end{figure*}
\end{widetext}

We suggest an interpretation of Eq.(\ref{flavordecomp_orb_proton}) 
and Eq.(\ref{flavordecomp_orb_Delta}) in terms of the geometric 
shapes of these baryons 
as depicted in Fig.~\ref{figure:shapes}.
Previously, by studying the electromagnetic $p\to \Delta^+$ 
transition in various baryon structure models, we have found 
that the proton has a positive intrinsic quadrupole moment $Q_0(p)$
corresponding to a prolate intrinsic charge distribution whereas the 
$\Delta^+$ has a 
negative intrinsic quadrupole moment of similar magnitude $Q_0(\Delta^+) 
\approx -Q_0(p)$  
corresponding to an oblate charge distribution~\cite{Hen08}. 
This appears to be 
consistent with our present findings for the quark orbital angular
momenta $L^{u}_z$ and $L^{d}_z$ in both systems.

\section{Summary}
\label{sec:5}
In summary, using a broken spin-flavor symmetry 
based parametrization of QCD, 
we have presented a straightforward calculation of the
quark spin and orbital angular momentum contributions to the total baryon spin.
For flavor octet baryons, we have shown that three-quark
operators reduce the standard quark model prediction based on
one-quark operators from $\Sigma_1 =1$ to
$\Sigma_1 = 0.41(12)$ in agreement with the experimental result.
On the other hand, in the case of flavor decuplet baryons, three-quark
operators enhance the contribution of one quark operators from
$\Sigma_3=3$ to $\Sigma_3=3.87(22)$. 

Assuming that the gluon contribution to baryon spin is small, 
we have suggested a qualitative interpretation of the positive 
and large $u$ quark and small 
$d$ quark orbital angular momenta in the proton in terms of a prolate  
quark distribution corresponding to a positive intrinsic quadrupole moment. 
In the case of the $\Delta^+$, $u$ and $d$ quarks have negative orbital 
angular momenta 
of the same magnitude corresponding to an oblate quark distribution 
giving rise to a negative intrinsic quadrupole moment. 
 



\end{document}